\documentclass[5p,twocolumn]{elsarticle}
\usepackage{graphicx,latexsym}
\usepackage{dcolumn}
\usepackage{amssymb,amsmath,bm}
\usepackage{subfigure}
\usepackage{braket}
\usepackage{siunitx}
\usepackage{array}
\usepackage{float}
\usepackage[nopar]{lipsum}
\usepackage{caption}
\usepackage{multirow}
\usepackage{bigstrut}

\usepackage{booktabs}

\biboptions{sort&compress}

\usepackage[super]{nth}

\newcommand{\angstrom}{\text{\normalfont\AA}}
\usepackage{hyperref}
\hypersetup{
    pdfnewwindow=true,       
    colorlinks=true,         
    linkcolor=blue,          
    citecolor=blue,          
    filecolor=magenta,       
    urlcolor=black           
}

\usepackage[normalem]{ulem}

\def\sec#1{Sec.\ \ref{#1}}

\def\fig#1{Fig.\ \ref{#1}}
\def\tab#1{Tab.\ \ref{#1}}

\journal{}

\begin{document}

\begin{frontmatter}


\title{Study of the buckling effects on the electrical and
	optical properties of the group III-Nitride monolayers}


\author[a1,a2]{Nzar Rauf Abdullah}
\ead{nzar.r.abdullah@gmail.com}
\address[a1]{Division of Computational Nanoscience, Physics Department, College of Science,
             \\ University of Sulaimani, Sulaimani 46001, Kurdistan Region, Iraq}
\address[a2]{Computer Engineering Department, College of Engineering,
	\\ Komar University of Science and Technology, Sulaimani 46001, Kurdistan Region, Iraq}

\author[a3]{Botan Jawdat Abdullah}
\address[a3]{Physics Department, College of Science- Salahaddin University-Erbil, Erbil 44001, Kurdistan Region, Iraq}

\author[a1]{Hunar Omar Rashid}

\author[a4]{Chi-Shung Tang}
\address[a4]{Department of Mechanical Engineering,
	National United University, 1, Lienda, Miaoli 36003, Taiwan}

\author[a5]{Vidar Gudmundsson}
\ead{vidar@hi.is}
\address[a5]{Science Institute, University of Iceland, Dunhaga 3, IS-107 Reykjavik, Iceland}


\begin{abstract}

We consider electronic and optical properties of group III-Nitride monolayers using first-principle calculations. The group III-Nitride monolayers have flat hexagonal structures with almost zero planar buckling, $\Delta$. By tuning the $\Delta$, the strong $\sigma\text{-}\sigma$ bond through sp$^2$ hybridization of a flat form of these monolayers can be changed to a stronger
$\sigma\text{-}\pi$ bond through sp$^3$ hybridization. Consequently, the band gaps of the monolayers are tuned due to a dislocation of the $s$- and $p$-orbitals towards the Fermi energy.
The band gaps decrease with increasing $\Delta$ for those flat monolayers, which have a band gap greater than $1.0$~eV, while no noticeable change or a flat dispersion of the band gap is seen for the flat monolayers, that have a band gap less than $1.0$~eV.
The decreased band gap causes a decrease in the excitation energy, and thus the static dielectric function, refractive index, and the optical conductivity are increased.
In contrast, the flat band gap dispersion of few monolayers in the group III-Nitride induces a reduction in the static dielectric function, the refractive index, and the optical conductivity.
We therefore confirm that tuning of the planar buckling can be used to control the physical properties of these monolayers, both for an enhancement and a reduction of the optical properties. These results are of interest for the design of optoelectric devices in nanoscale systems.

\end{abstract}


\end{frontmatter}

\section{Introduction}
Two-dimensional (2D) materials, which are crystalline solids with just one or few atomic layers, have been a significant topic in recent years due to their unusual characteristics \cite{butler2013progress, bhimanapati2015recent, abdullah2021properties}. A rising variety of 2D materials have been investigated in models and experiments since the discovery of graphene \cite{novoselov2004electric, ABDULLAH2022115554}. Many of the 2D materials experimentally investigated have exhibited remarkable new features not found in bulk materials \cite{vogt2012silicene, ABDULLAH2021106981, mak2016photonics}. It is essential to consider how such promising properties of the 2D materials have come to be applied in both theoretical and practical applications \cite{ABDULLAH2022106409}.

Monolayer materials with fascinating features, including group III-Nitride monolayers, such as BN, AlN, GaN, InN, and TlN, exist in addition to graphene and are valuable in a variety of applications. The BN monolayer has been synthesized efficaciously using a variety of techniques, making it one of the most well-known 2D materials \cite{han2008structure, stehle2015synthesis}. The AlN monolayer has recently been synthesized using plasma assisted molecular beam epitaxy \cite{tsipas2013evidence}. The GaN monolayer was synthesized using a variety of techniques \cite{seo2015direct, al2016two}. Several InN nanostructures, such as InN nanotubes, nanowires \cite{xu2007synthesis}, and monolayer InN quantum wells, have also been produced \cite{yoshikawa2008fabrication}. Although no TlN monolayer has been synthesized, a theoretical examination using DFT revealed that it has the same structure as other metallic nitrides, implying that it may be produced \cite{shi2010structural, elahi2016investigation}.

Doping, strain, electric fields, heterostructure, chemical functionality, and planar buckling tuning are all a several of the techniques used to change the physical properties of monolayer materials. The band gap is likely to vary if the electrical structure of systems changes, and doping is one of the strategies used to influence the physical characteristics of 2D monolayers. The electronic gap in III-Nitride monolayers is reduced due to the size and symmetry of the doping atoms, resulting in a considerable reduction in energy gaps as well as a possible change in the optical properties, depending on the type of doping atom \cite{oliveira2019electronic, javaheri2018electronic, alaal2018tuning, ABDULLAH2021110095, das2021electronic}.

Another factor is strain, which is investigated using ab-initio calculations to explore its effects  son the electrical properties of III-nitride monolayers, and the findings show that the free-strains of BN, AlN, GaN, and InN monolayers result in an indirect band gap. The band gap in GaN and InN is changed from indirect to direct with compressive biaxial strain, but the band gap in BN and AlN is unchanged. Only the BN monolayer acts as a direct band gap semiconductor when subjected to tensile stress \cite{ghasemzadeh2018strain}. Strain and stress are employed to adjust the electrical and optical properties of single-layer boron nitride using full potential augmented plane waves plus local orbitals in the context of density functional theory. The findings show that adding stress and strain to a BN nanosheet may change its energy band gap. Stress and strain have minimal influence on the reflectivity spectrum while shifting the energy gap direction, according to optical simulations \cite{jalilian2016tuning}.

Both AlN and GaN have an indirect band gap at zero strain, however strain has an influence on the electronic characteristics of AlN and GaN, and the gap lowers for tensile uniform strain and rises for compressive uniform strain. Furthermore, AlN remains an indirect gap material for all strain values studied, whereas GaN transitions from an indirect to a direct gap at compressive strain \cite{postorino2020strain}, and the InN monolayer is found to be a direct band gap semiconductor in the region of 6 applied biaxial strain. The energy band gap of the materials under consideration is also reduced when stress and strain are applied. The optical calculations show that adding stress and strain on a system causes the optical spectra to shift blue and red \cite{jalilian2016tuning}.

Another factor that influences the physical properties of 2D monolayers is an electric field; for example, it is possible to minimize the electrical band gap of BN monolayers by combining P and As substitute for N doping with an external electric field. Electric fields ranging from $\text{-}0.5$ to 0.5 (eV/e) perpendicular to the BN monolayer are applied in both directions along the z-axis. When subjected to a 0.1 (eV/e) electric field, the electronic band gap did not noticeably change. On the other hand, a stronger electric field can considerably reduce this property \cite{hoat2020reducing, ABDULLAH2021114644}.

A further factor that affects the material characteristics of 2D monolayers is a placement in a van der Waals hetrostructure. According to the optical spectra, graphene monolayers stacked with BN monolayers  widen the band gap of grapheme, and absorb light throughout a wide frequency range, from near IR to UV \cite{aggoune2020structural}.

Other interesting strategy for altering the material characteristics of 2D monolayers is an approach to tune the planar buckling. Previously, the effects of buckling on the electrical and optical characteristics of beryllium oxide monolayers were investigated using a first-principles approach with a considerable level of planar buckling. According to electrical investigations, planar buckling can result in a tunable band gap since the energy band gap decreases when the planar buckling parameter is increased. Furthermore, as the planar buckling factor is increased, all optical spectra red-shift to lower energy. These changes are produced by planar buckling weakening bonds and increasing the number of bonds \cite{jalilian2016buckling}. It has been proposed that the buckling of GaAs monolayer can be tuned by applying an external electric field \cite{doi:10.1063/1.4979507}.

In this work, a first-principle density functional theory is used to investigate the electrical and optical properties of two-dimensional BN, AlN, GaN, InN, and TlN structures. The effect of planar buckling on the band structure, which is essential in the electrical system, are studied.
The band gap of BN, AlN, and GaN varies significantly when the planar buckling is increased, although the effects of change are smaller for both InN and TlN, which affects the optical characteristics such as the excitation energy, the dielectric function, the refractive index, and the optical conductivity.

The paper is organized as follow: The computational methodologies and model structure are briefly shown in \sec{Sec:Methodology}. The major achieved are examined in \sec{Sec:Results}. The conclusion of the results is presented in \sec{Sec:Conclusion}

\section{Computational Tools}\label{Sec:Methodology}
Our calculations are performed within the first principles via the density functional theory, DFT, as implemented in Quantum espresso software \cite{Giannozzi_2009, giannozzi2017advanced}.
The generalized gradient approximation (GGA) in the Perdew-Burke- Ernzernhof (PBE) functionals is used for calculating the band structure, the density of states, and the optical properties of all considered monolayers \cite{PhysRevLett.77.3865, doi:10.1063/1.1926272, ABDULLAH2022106835}.
A vacuum space is considered to be $20 \, \angstrom$ to exclude the interaction between the nanosheets.
In the structure relaxation, a $18\times18\times1$ Monkhorst-Pack K-point grid, and an energy cutoff $1088$~eV are used. In order to get the fully relaxed monolayer structures, forces are assumed to be smaller than $0.001$ eV/$\angstrom$.
In addition, a K-point $18\times18\times1$-grid for self-consistency, and a $100\times100\times1$ grid for the DOS calculations are used with the same energy cutoff mentioned above. Finally, the optical broadening is $0.1$~eV in the calculations \cite{ABDULLAH2022114590}.

\section{Results}\label{Sec:Results}

\begin{figure}[htb]
	\centering
	\includegraphics[width=0.45\textwidth]{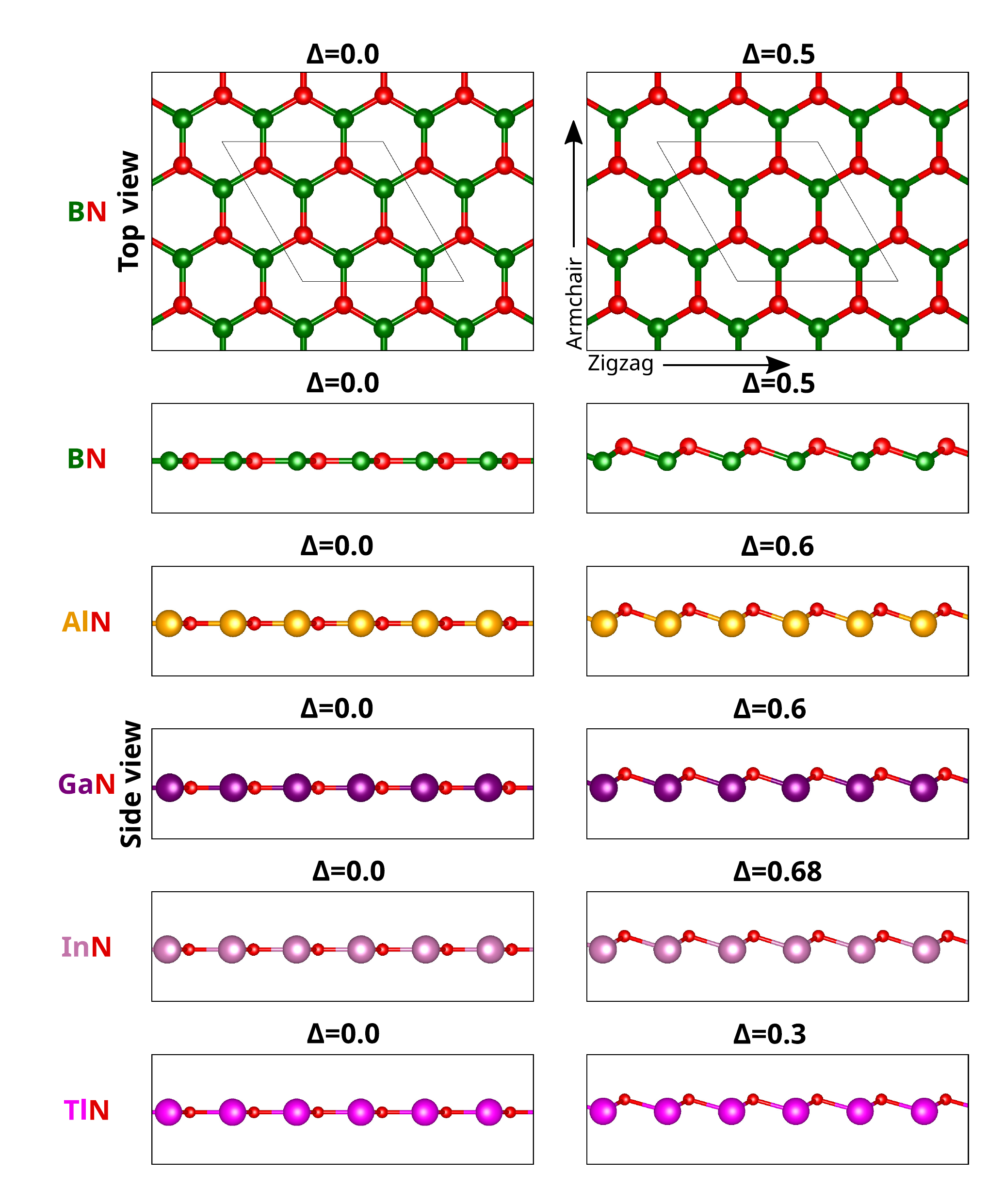}
	\caption{Crystal structures of BN, AlN, GaN, InN, and TlN for $\Delta = 0.0$ (left panel) and the maximum allowed value of $\Delta$ (right panel) for each monolayer.
	The top most figures display the BN monolayer in top views, and other figures represent side views of each of the considered monolayer.}
	\label{fig01}
\end{figure}

In this section, the electronic and the optical properties of group III-Nitride monolayers are investigated by considering different values of planar buckling, where the monolayers are in $2\times2\times1$ supercells. The crystal structures of all the considered monolayers are shown in \fig{fig01}. The monolayers are presented for $\Delta = 0.0$ (left panel), and for the the maximum allowed value of $\Delta$ (right panel) for each monolayers. In a flat or planar group III-Nitride monolayer ($\Delta = 0.0$), all the N and the group III atoms are situated in the same $xy$-plane as it is presented in \fig{fig01} (left panel).
If a planar buckling is considered ($\Delta \neq 0.0$), the group III atoms are located in the same plane and all the N atoms are situated in another plane. The planar bucking indicates the vertical distance, $\Delta$, between the group III atoms and the N planes.

We first present the formation energy of the monolayers, which is the energy needed for forming the atomic configuration of a structure. The formation energy of planar or flat group III-Nitride monolayers, $\Delta = 0.0$, is $\text{-}13.33$ (BN), $\text{-}10.16$ (AlN), $\text{-}7.59$ (GaN), $\text{-}6.05$ (InN), and $\text{-}4.40$~eV (TlN), where $\Delta$ is the buckling constant or the planar buckling that represents the degree of the buckling.
A more negative formation energy is obtained for a more favored stable monolayer indicating that
a smaller atom (B and N atoms) is beneficial to the formation of a planar structure and thus also to a layered 2D structure. So, the BN monolayer is the most electronically stable structure among all the considered monolayers.
We stress that the flat form of these monolayers does not have precisely zero value for the buckling constant, but there is a tiny buckling due to the differences of the atomic diameter. For example, the buckling parameter is $0.001$~$\angstrom$ for the first four flat monolayers from BN to InN,
while it is $0.061$~$\angstrom$ for flat TlN.

If the planar buckling is increased, the lattice constants, $a$, are unchanged while the bond lengths of the monolayers increase. The values of the lattice constant for BN, AlN, GaN, InN, and TlN monolayers are $2.507$, $3.117$, $3.213$, $3.553$, and $3.719$~$\angstrom$, respectively, for all values of planar buckling.
The bond lengths for different values of planar buckling are presented in \tab{table_one} for all the considered monolayers, and the bond lengths at $\Delta=0.0$ agree very well with previous DFT calculations \cite{Liu_2019}.
The longer bond lengths mean the electrons are less tightly bound to the atoms, and hence require less energy to be removed, leading to a decreased band gap \cite{gacevic:hal-00632224}. Consequently, the longer bond lengths corresponds to smaller band gaps. So, one may expect that the TlN monolayer has the smallest band gap among all the considered monolayers because it has the longest bond length even at $\Delta = 0.0$.

\begin{table}[!h]
			\centering
	\begin{tabular}{ l l l l l l }
		\cline{1-1}
\hline
$\Delta$ ($\angstrom$)& BN & AlN   & GaN   &  InN  & TlN   \\
\hline
0.0 & 1.447  & 1.799 & 1.885 & 2.051 & 2.147  \\
0.1 & 1.451  & 1.802 & 1.858 & 2.054 & 2.156 \\
0.2 & 1.461  & 1.810 & 1.866 & 2.061 & 2.184 \\
0.3 & 1.478  & 1.824 & 1.879 & 2.073 & 2.229  \\
0.4 & 1.501  & 1.843 & 1.898 & 2.090 & --     \\
0.5 & 1.531  & 1.868 & 1.921 & 2.111 & --     \\
0.6 &  ---   & 1.897 & 1.950 & 2.137 & --     \\
\hline
	\end{tabular}
\caption{The bond length, $l$, in the case of different values of planar buckling for all considered monolayers.}
\label{table_one}
\end{table}

\begin{figure}[htb]
	\centering
	\includegraphics[width=0.45\textwidth]{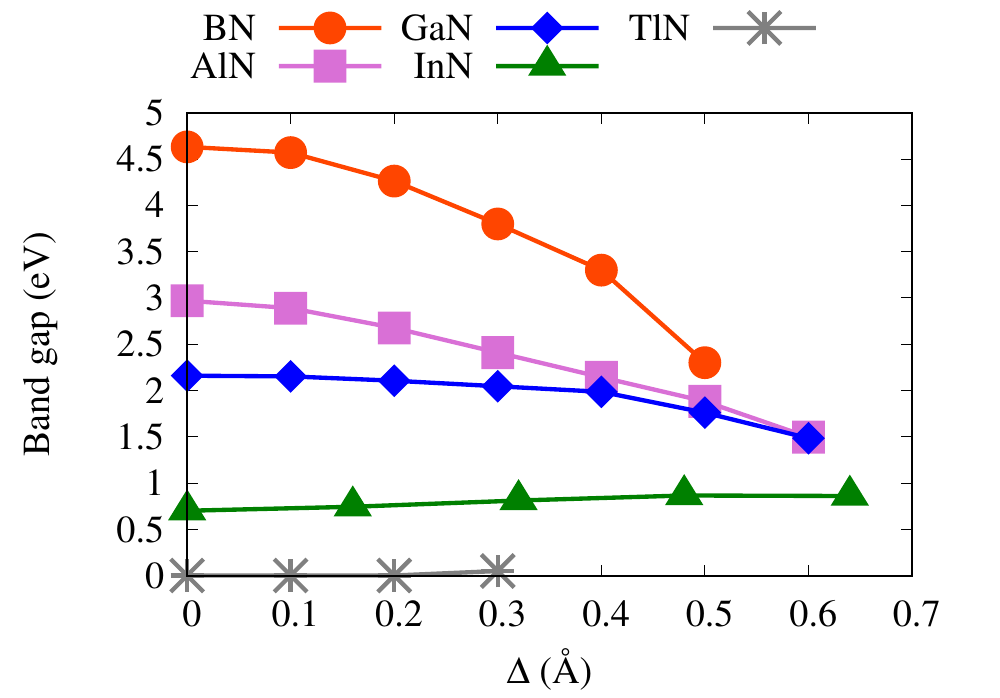}
	\caption{Band gap of BN (red), AlN (pink), GaN (blue), InN (green), and TlN (gray) as a function of planar buckling, $\Delta$.}
	\label{fig02}
\end{figure}

The band gap of the monolayers as a function of the planer buckling is plotted in \fig{fig02}.
We first explain the $\Delta$ limitation values before describing the band gap behavior with increasing planar buckling.
The critical values of the planar buckling depend on the hybridization between the $s$- and the $p$-orbitals.
The maximum allowed value of $\Delta$ is found by the $s$- and the $p$-orbital hybridization.
The link between the hybridization and the bond angle for $s\text{-}p$ hybrids is determined by this equation,
$\cos(\theta) = s/(s-1) = (p-1)/p$, where $\theta$ is the angle between the respective orbitals and the $s$ and the $p$ parameters are given as decimal fractions \cite{kaufman1993inorganic}.
All the considered monolayers here have a $sp^2$-hybridization when $\Delta = 0.0$, and the $sp^2$ is changed to a $sp^3$-hybridization when the planar buckling is increased \cite{jalilian2016buckling}. The values obtained for $\Delta$ are acceptable if the hybridization between the $s$- and the $p$-orbitals remains steady between the $sp^2$ and the $sp^3$. We therefore see that the maximum allowed value of $\Delta$ is not the same for all these monolayers.

The band gap of the flat BN, AlN, GaN, InN, and TlN monolayers, at $\Delta = 0.0$, are
$4.63$, $2.97$, $2.16$, $0.702$, and $0.002$~eV, respectively. These values for the band gap are in agreement with previous DFT calculations \cite{Liu_2019}.
The band gap is the largest for the BN and the smallest for the TlN monolayer at $\Delta = 0.0$, which can be referred back to the shortest B-N bond length and the longest Tl-N bond length.
Furthermore, the band gap is indirect for flat AlN and GaN monolayers while it is direct for flat BN, InN, and TlN monolayers.
We should remember that a hybrid density functional proposed by Heyd, Scuseria, and Ernzerhof (HSE) would be a better choice for calculating the band gap \cite{doi:10.1063/1.1564060}. The band gap of the flat BN, AlN, GaN, InN, and TlN monolayers using the HSE are $5.67$, $4.04$, $3.43$, $1.61$, and $0.14$~eV, respectively. So, one can see that the band gap of these monolayers is underestimated using the GGA-PBE method. However, we still use the standard GGA-PBE approach to carry out most of our calculations here.

\begin{figure}[htb]
	\centering
	\includegraphics[width=0.5\textwidth]{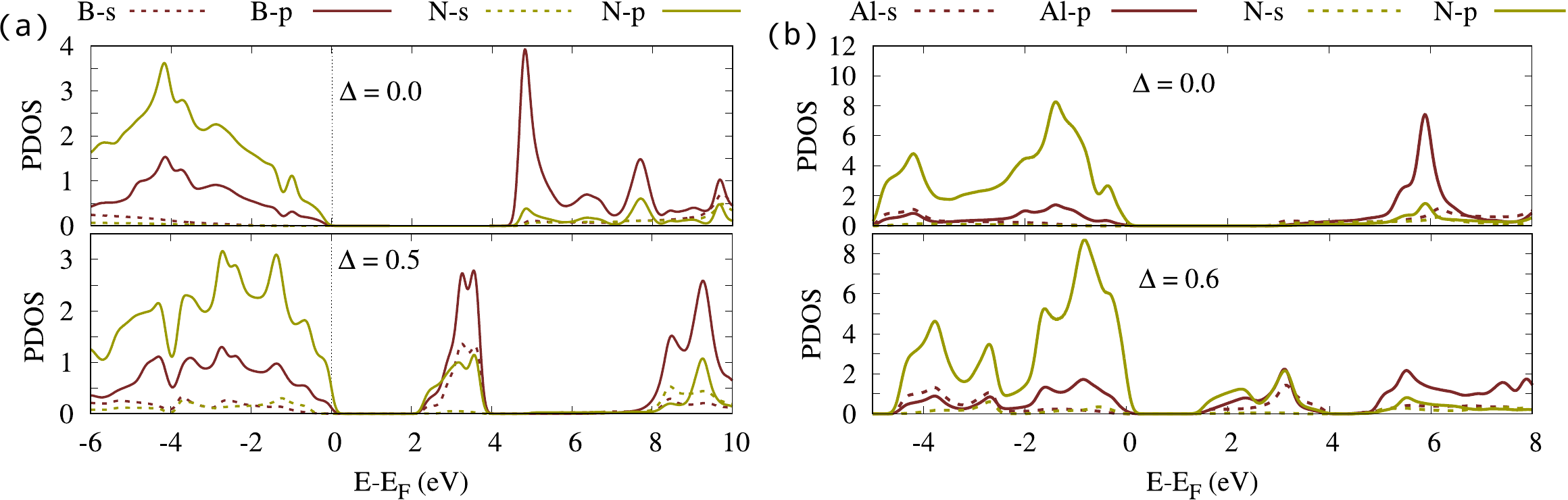}
	\caption{Partial density of state of BN (a), and AlN (b) monolayers for $\Delta = 0.0$ and maximum allowed value of $\Delta$.}
	\label{Pdos}
\end{figure}

Once all atoms are situated in the same plane ($\Delta = 0.0$),
strong $\sigma$ bonds are formed through $sp^2$ orbital overlapping.
This orbital contribution can be seen in the density of state shown in \fig{Pdos}, where 
the partial density of state of two of these monolayers are plotted such as BN and AlN, and other monolayers have the same characteristics.

If the planar buckling is increased, the planar buckling reduces the $sp^2$
overlapping and changes it towards to $sp^3$. As the results, the bond lengths of the monolayers are increased (see \tab{table_one}) and thus the bond symmetry is broken simultaneously.
The increased rate of the bond length for the BN, AlN, and GaN monolayers is higher than that of the InN and TlN monolayers. As a result, the band gap is noticeably decreased for the BN, AlN, and GaN, while the almost flat dispersion for the InN and TlN monolayers is noticed with a slightly increased band gap of InN at the higher values of $\Delta$.

In fact, not only the band gap is tuned by increasing the planar buckling, but the whole band structure of these monolayers is modified.
Increasing the planar buckling causes a deformation of the structures of these monolayers, which leads to a symmetry breaking of the hexagonal structure.
The monolayers having a shorter bond length display a stronger symmetry breaking, when the planar buckling is increased. In this case, the band gap reduction arises via the breaking of the $sp^2$ bond symmetry of these monolayers. The structures change to $sp^3$ bonds in which the $s$- and the the $p$-orbital of the atoms move towards the Fermi energy resulting in a reduction of the semiconducting energy band gap. In addition, the indirect band gaps of the AlN and GaN monolayers transform to direct band gaps at higher values of $\Delta$, and the direct band gaps of BN, InN, and TlN monolayer keep their character of direct band gaps.
We have previously reported this for a BN monolayer \cite{abdullah2022enhanced}.
The final remarkable point about the band gaps is that the band gap of the AlN and GaN monolayers at $\Delta = 0.4$, $0.5$, and $0.6$~$\angstrom$ almost coincide. This is caused by the similar lengths of the Ga-N and the Al-N bonds obtained for these values of $\Delta$.

\begin{figure}[htb]
	\centering
	\includegraphics[width=0.45\textwidth]{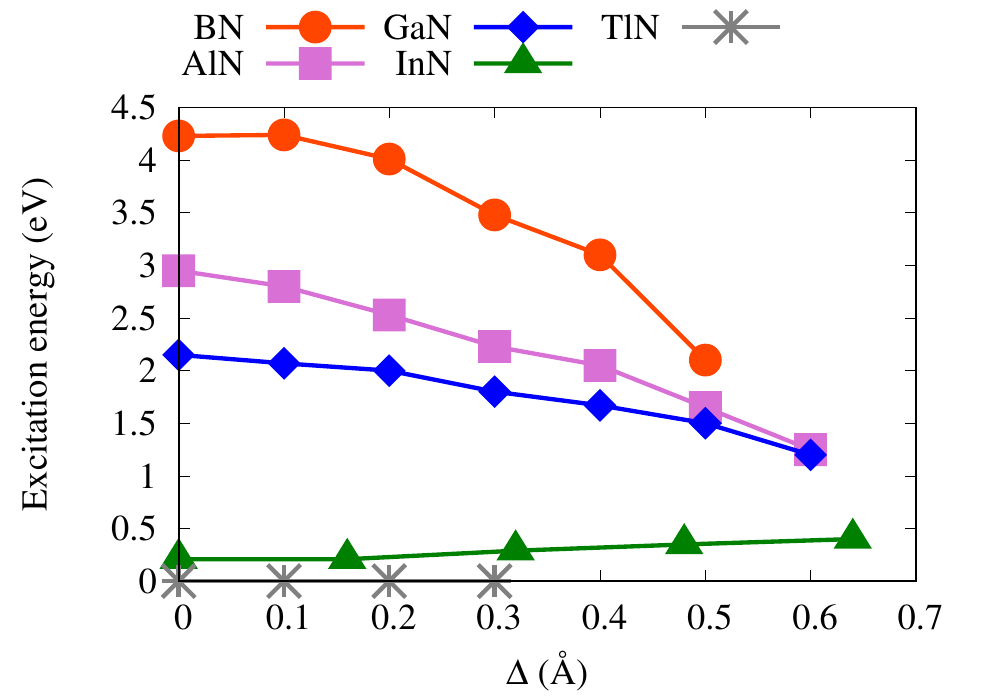}
	\caption{Excitation energy spectra of BN (red), AlN (pink), GaN (blue), InN (green), and TlN (gray) as a function of planar buckling, $\Delta$ for  E$_{\rm \parallel}$.}
	\label{fig03}
\end{figure}

It is interesting to see the influences of the planar buckling on the optical properties of the monolayers. We have used the random phase approximation (RPA) \cite{Ren2012} implemented in the QE software package, in which the excitation energy, the static dielectric function, the refractive index, and the optical conductivity are calculated.
The excitation energy spectra versus planar buckling is presented in \fig{fig03}. The excitation energy is the minimum energy required to transfer an electron from the valence to the conduction bands. In the calculations of the optical properties, we consider the polarization of the incident light parallel to the surface of the monolayers, E$_{\parallel}$, and we can confirm that the results for perpendicular polarized light, E$_{\perp}$, are qualitatively similar to the results for E$_{\parallel}$. We thus present only the results for E$_{\parallel}$.

The trends of the excitation energy spectra follow the band gap spectra and we can see that the flat BN monolayer
has maximum excitation spectra among all the considered structures at $\Delta = 0.0$ as the BN monolayer has the largest band gap. The excitation energy is decreased with increasing value of $\Delta$ for the BN, AlN, and GaN monolayers because the band gaps of these monolayers decrease and a direct band gap is seen at high values of $\Delta$, indicating that less energy is required to transfer the electrons to the conduction band from the valence band region.

\begin{figure}[htb]
	\centering
	\includegraphics[width=0.4\textwidth]{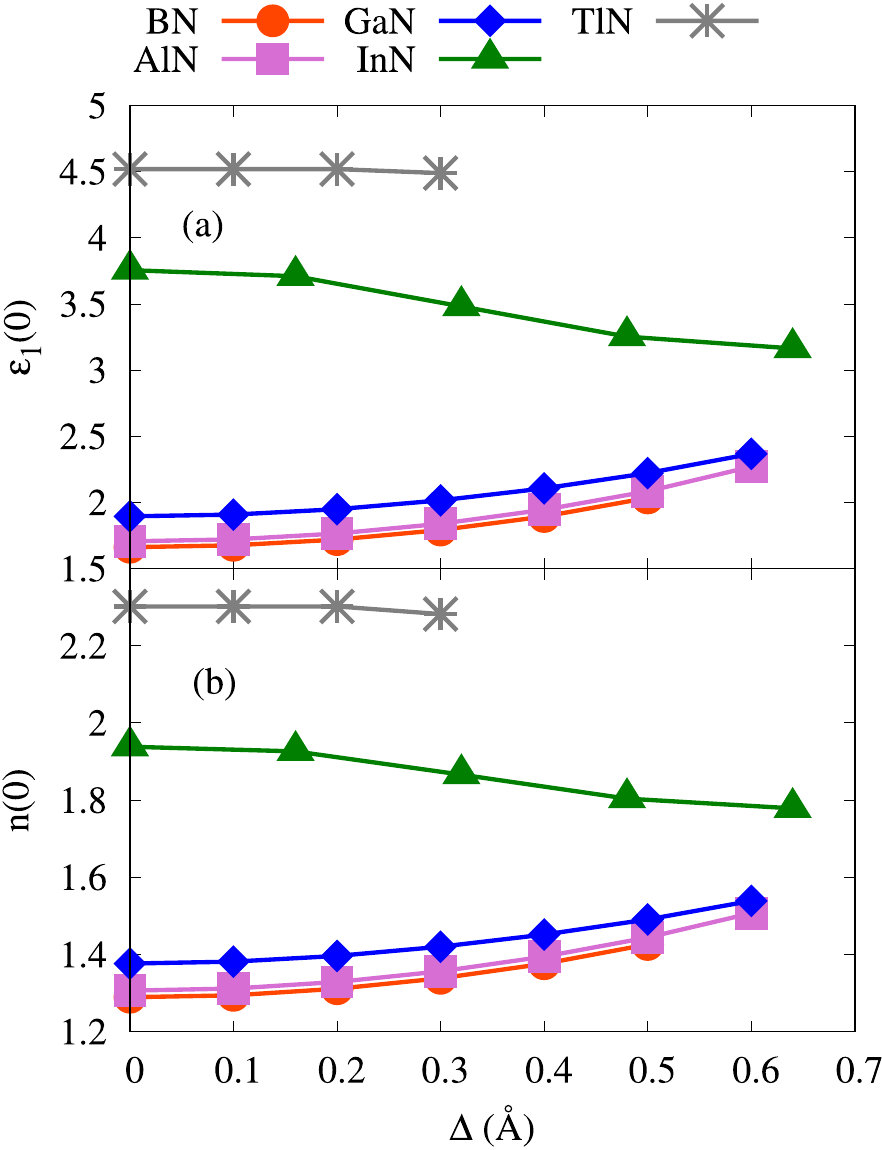}
	\caption{(a) Static dielectric function, $\varepsilon_1(0)$, and (b) refractive index, $n(0)$, of BN (red), AlN (pink), GaN (blue), InN (green), and TlN (gray) as a function of planar buckling, $\Delta$ for  E$_{\rm \parallel}$.}
	\label{fig04}
\end{figure}

We next show the effects of planar buckling on the dielectric function and the refractive index of the monolayers. The static dielectric function is the real part of dielectric function, $\varepsilon_1(\omega)$, at zero energy value for the incident light, $\varepsilon_1(0)$.
The $\varepsilon_1(0)$ (a) and the corresponding refractive index, $n(0)$, (b) are shown in \fig{fig04}. The $\varepsilon_1(\omega)$ introduces a material's ability to interact with an electric field (store and remit energy) without absorbing energy.
In fact, the value of $\varepsilon_1(0)$ is strongly related to the band gaps of the monolayers. It has been demonstrated that $\varepsilon_1(0)$ is inversely proportional to the band gap $\varepsilon_1(0) \approx 1/E_{g}$ \cite{PhysRev.128.2093}.
We therefore see the static dielectric function is increased for the BN, AlN, and GaN monolayers, when the planar buckling is increased, while it is decreased for the InN monolayer and has an almost flat dispersion for the TlN monolayer. Additionally, the TlN monolayer has the highest values of $\varepsilon_1(0)$ and $n(0)$ among the monolayers throughout all values of the planar buckling. This should be expected as it has the smallest band gap, and the band gap is almost unchanged with planar buckling.
It indicates that the TlN monolayer has the highest ability to store energy throughout all values of planar buckling.

Additionally, the $n(\omega)$ spectra have almost the same qualitative characteristics as $\varepsilon_1(\omega)$ for all the monolayers, and we see that the $n(0)$ spectra are directly proportional to the $\varepsilon_1(0)$ spectra. These phenomena are caused by the variation of the in-plane bond nature, in which the $\sigma\text{-}\sigma$ bonds become weaker and the $\sigma$-bonds move towards a side-by-side overlapping as a result of the increased planar buckling.

\begin{figure}[htb]
	\centering
	\includegraphics[width=0.45\textwidth]{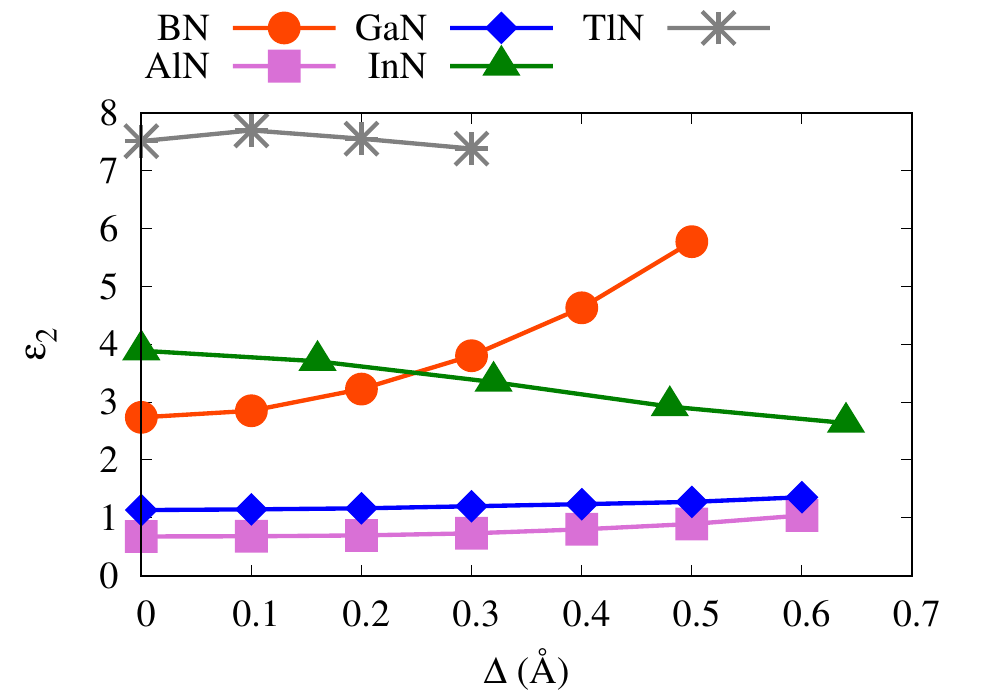}
	\caption{Imaginary part of the dielectric function, $\varepsilon_2$, of BN (red), AlN (pink), GaN (blue), InN (green), and TlN (gray) as a function of planar buckling, $\Delta$ for  E$_{\rm \parallel}$. }
	\label{fig05}
\end{figure}

Another interesting property of the monolayers is the imaginary part of dielectric function, $\varepsilon_2$. The imaginary part of the dielectric constant introduces a material's ability to permanently absorb energy from a time-varying electric field. The $\varepsilon_2$ as a function of the planar buckling is plotted in \fig{fig05}. The $\varepsilon_2$ here is the value of the first intense peak corresponding to the electron transition in the optical band gap.
The TlN monolayer has the highest value of $\varepsilon_2$, which is due to the smallest band gap among the considered monolayers, and the $\varepsilon_2$ value is almost constant because the band gap is almost unchanged when the $\Delta$ is increased. It indicates that the TlN monolayer has a high ability to absorb energy from the electric field, and this ability is not changed by tuning the $\Delta$. Followed by TlN monolayer, the $\varepsilon_2$ value of the InN monolayer is also higher than that of the BN, AlN, and GaN monolayers, and it has a flat dispersion for low values of $\Delta$, but it  decreases a bit for the high values of $\Delta$.
Furthermore, the $\varepsilon_2$ values of the BN, AlN, and GaN monolayers are increased with planar buckling, especially, there is a pronounced enhancement of  $\varepsilon_2$ for the BN monolayer.
This demonstrates that the ability of energy absorption of these three monolayers is increased with increasing $\Delta$.

\begin{figure}[htb]
	\centering
	\includegraphics[width=0.45\textwidth]{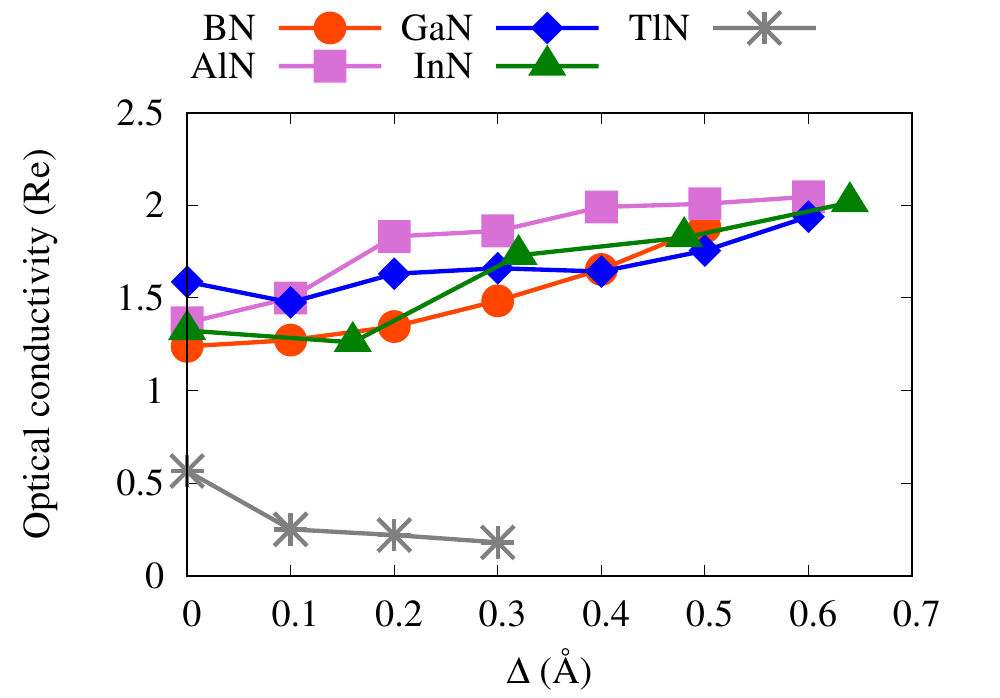}
	\caption{Optical conductivity, real part, of BN (red), AlN (pink), GaN (blue), InN (green), and TlN (gray) as a function of planar buckling, $\Delta$ for  E$_{\rm \parallel}$.  }
	\label{fig06}
\end{figure}

Finally, we present the influences of the planar buckling on the optical conductivity. The real part of the optical conductivity versus the planar buckling is presented in \fig{fig06}. The optical conductivity follows the characteristics of $\varepsilon_2$. It can be seen that the optical conductivity is enhanced for the BN, AlN, GaN and InN monolayers, when the planar buckling is increased, while a slight reduction for TlN is seen.
The reduction of the optical conductivity for TlN with planar buckling is expected as $\varepsilon_2$ is also reduced by increased with the planar buckling. In contrast, a gradual enhancement of the optical conductivity of a BN monolayer is seen. So, we can clearly observe that the effect of planar buckling on the electronic and the optical properties of the group III-Nitride monolayers is not the same. In some monolayers, an enhancement in the optical properties is recorded, while in some other monolayers a reduction is seen, when the planar buckling is increased.

\section{Conclusion}\label{Sec:Conclusion}

We have used density functional theory formalism to investigate the electronic and the optical properties of the group III-Nitride monolayers, where the influences of a planar buckling plays an essential role.
The band gap of the monolayers is calculated using both the GGA-PBE and the HSE functionals. The increase in the planar buckling causes a decrease in the band gap of the monolayers, that have short bond lengths, but an unchanged band gap of those monolayers that have longer bond lengths. This is because the planar buckling breaks the hexagonal symmetry of the monolayer, and the symmetry of the monolayers with short bond lengths is strongly affected compared to the monolayers with long bond lengths.
Furthermore, the tuned band gaps directly control the optical properties, such as the static dielectric functions, the refractive index, and the optical conductivity. We see an enhancement of the optical properties from the Deep-UV region to the visible light range, which will be very useful for optical communication devices.

\section{Acknowledgment}
This work was financially supported by the University of Sulaimani and
the Research center of Komar University of Science and Technology.
The computations were performed on resources provided by the Division of Computational
Nanoscience at the University of Sulaimani.



\begin{thebibliography}{10}
	\expandafter\ifx\csname url\endcsname\relax
	\def\url#1{\texttt{#1}}\fi
	\expandafter\ifx\csname urlprefix\endcsname\relax\def\urlprefix{URL }\fi
	\expandafter\ifx\csname href\endcsname\relax
	\def\href#1#2{#2} \def\path#1{#1}\fi
	
	\bibitem{butler2013progress}
	S.~Z. Butler, S.~M. Hollen, L.~Cao, Y.~Cui, J.~A. Gupta, H.~R. Guti{\'e}rrez,
	T.~F. Heinz, S.~S. Hong, J.~Huang, A.~F. Ismach, et~al., Progress,
	challenges, and opportunities in two-dimensional materials beyond graphene,
	ACS nano 7~(4) (2013) 2898--2926.
	
	\bibitem{bhimanapati2015recent}
	G.~R. Bhimanapati, Z.~Lin, V.~Meunier, Y.~Jung, J.~Cha, S.~Das, D.~Xiao,
	Y.~Son, M.~S. Strano, V.~R. Cooper, et~al., Recent advances in
	two-dimensional materials beyond graphene, ACS nano 9~(12) (2015)
	11509--11539.
	
	\bibitem{abdullah2021properties}
	N.~R. Abdullah, B.~J. Abdullah, C.-S. Tang, V.~Gudmundsson, Properties of BC$_6$N
	monolayer derived by first-principle computation: Influences of interactions
	between dopant atoms on thermoelectric and optical properties, Materials
	Science in Semiconductor Processing 135 (2021) 106073.
	
	\bibitem{novoselov2004electric}
	K.~S. Novoselov, A.~K. Geim, S.~V. Morozov, D.-e. Jiang, Y.~Zhang, S.~V.
	Dubonos, I.~V. Grigorieva, A.~A. Firsov, Electric field effect in atomically
	thin carbon films, Science 306~(5696) (2004) 666--669.
	
	\bibitem{ABDULLAH2022115554}
	N.~R. Abdullah, H.~O. Rashid, C.-S. Tang, A.~Manolescu, V.~Gudmundsson,
	\href{https://www.sciencedirect.com/science/article/pii/S0921510721005080}{Controlling
		physical properties of bilayer graphene by stacking orientation caused by
		interaction between B and N dopant atoms}, Materials Science and Engineering:
	B 276 (2022) 115554.
	\newblock \href {https://doi.org/https://doi.org/10.1016/j.mseb.2021.115554}
	{\path{doi:https://doi.org/10.1016/j.mseb.2021.115554}}.
	\newline\urlprefix\url{https://www.sciencedirect.com/science/article/pii/S0921510721005080}
	
	\bibitem{vogt2012silicene}
	P.~Vogt, P.~De~Padova, C.~Quaresima, J.~Avila, E.~Frantzeskakis, M.~C. Asensio,
	A.~Resta, B.~Ealet, G.~Le~Lay, Silicene: compelling experimental evidence for
	graphenelike two-dimensional silicon, Physical review letters 108~(15) (2012)
	155501.
	
	\bibitem{ABDULLAH2021106981}
	N.~R. Abdullah, H.~O. Rashid, V.~Gudmundsson,
	\href{https://www.sciencedirect.com/science/article/pii/S0749603621001798}{Study
		of BC$_{14}$N-bilayer graphene: Effects of atomic spacing and interatomic
		interaction between B and N atoms}, Superlattices and Microstructures 156
	(2021) 106981.
	\newblock \href {https://doi.org/https://doi.org/10.1016/j.spmi.2021.106981}
	{\path{doi:https://doi.org/10.1016/j.spmi.2021.106981}}.
	\newline\urlprefix\url{https://www.sciencedirect.com/science/article/pii/S0749603621001798}
	
	\bibitem{mak2016photonics}
	K.~F. Mak, J.~Shan, Photonics and optoelectronics of 2D semiconductor
	transition metal dichalcogenides, Nature Photonics 10~(4) (2016) 216--226.
	
	\bibitem{ABDULLAH2022106409}
	N.~R. Abdullah, B.~J. Abdullah, V.~Gudmundsson,
	\href{https://www.sciencedirect.com/science/article/pii/S136980012100740X}{High
		thermoelectric and optical conductivity driven by the interaction of Boron
		and Nitrogen dopant atoms with a 2D monolayer beryllium oxide}, Materials
	Science in Semiconductor Processing 141 (2022) 106409.
	\newblock \href {https://doi.org/https://doi.org/10.1016/j.mssp.2021.106409}
	{\path{doi:https://doi.org/10.1016/j.mssp.2021.106409}}.
	\newline\urlprefix\url{https://www.sciencedirect.com/science/article/pii/S136980012100740X}
	
	\bibitem{han2008structure}
	W.-Q. Han, L.~Wu, Y.~Zhu, K.~Watanabe, T.~Taniguchi, Structure of chemically
	derived mono-and few-atomic-layer Boron Nitride sheets, Applied Physics
	Letters 93~(22) (2008) 223103.
	
	\bibitem{stehle2015synthesis}
	Y.~Stehle, H.~M. Meyer~III, R.~R. Unocic, M.~Kidder, G.~Polizos, P.~G. Datskos,
	R.~Jackson, S.~N. Smirnov, I.~V. Vlassiouk, Synthesis of hexagonal Boron
	Nitride monolayer: control of nucleation and crystal morphology, Chemistry of
	materials 27~(23) (2015) 8041--8047.
	
	\bibitem{tsipas2013evidence}
	P.~Tsipas, S.~Kassavetis, D.~Tsoutsou, E.~Xenogiannopoulou, E.~Golias,
	S.~Giamini, C.~Grazianetti, D.~Chiappe, A.~Molle, M.~Fanciulli, et~al.,
	Evidence for graphite-like hexagonal AlN nanosheets epitaxially grown on
	single crystal ag (111), Applied Physics Letters 103~(25) (2013) 251605.
	
	\bibitem{seo2015direct}
	T.~H. Seo, A.~H. Park, S.~Park, Y.~H. Kim, G.~H. Lee, M.~J. Kim, M.~S. Jeong,
	Y.~H. Lee, Y.-B. Hahn, E.-K. Suh, Direct growth of GaN layer on carbon
	nanotube-graphene hybrid structure and its application for light emitting
	diodes, Scientific reports 5~(1) (2015) 1--7.
	
	\bibitem{al2016two}
	Z.~Y. Al~Balushi, K.~Wang, R.~K. Ghosh, R.~A. Vil{\'a}, S.~M. Eichfeld, J.~D.
	Caldwell, X.~Qin, Y.-C. Lin, P.~A. DeSario, G.~Stone, et~al., Two-dimensional
	Gallium Nitride realized via graphene encapsulation, Nature materials 15~(11)
	(2016) 1166--1171.
	
	\bibitem{xu2007synthesis}
	H.~Xu, Z.~Liu, X.~Zhang, S.~Hark, Synthesis and optical properties of InN
	nanowires and nanotubes, Applied physics letters 90~(11) (2007) 113105.
	
	\bibitem{yoshikawa2008fabrication}
	A.~Yoshikawa, S.~Che, N.~Hashimoto, H.~Saito, Y.~Ishitani, X.~Wang, Fabrication
	and characterization of novel monolayer inn quantum wells in a GaN matrix,
	Journal of Vacuum Science \& Technology B: Microelectronics and Nanometer
	Structures Processing, Measurement, and Phenomena 26~(4) (2008) 1551--1559.
	
	\bibitem{shi2010structural}
	L.~Shi, Y.~Duan, L.~Qin, Structural phase transition, electronic and elastic
	properties in TlX (X= N, P, As) compounds: Pressure-induced effects,
	Computational materials science 50~(1) (2010) 203--210.
	
	\bibitem{elahi2016investigation}
	S.~Elahi, M.~Farzan, H.~Salehi, M.~Abolhasani, An investigation of electronic
	and optical properties of TlN nanosheet and compare with TlN bulk (Wurtzite)
	by first principle, Optik 127~(20) (2016) 9367--9376.
	
	\bibitem{oliveira2019electronic}
	L.~M. Oliveira, O.~Santos, J.~Martins, S.~Azevedo, J.~Kaschny, Electronic and
	optical properties of Ge doped graphene and BN monolayers, Applied Physics A
	125~(11) (2019) 1--8.
	
	\bibitem{javaheri2018electronic}
	S.~Javaheri, M.~Babaeipour, A.~Boochani, S.~Naderi, Electronic and optical
	properties of V doped AlN nanosheet: DFT calculations, Chinese Journal of
	Physics 56~(6) (2018) 2698--2709.
	
	\bibitem{alaal2018tuning}
	N.~Alaal, I.~S. Roqan, Tuning the electronic properties of hexagonal
	two-dimensional GaN monolayers via doping for enhanced optoelectronic
	applications, ACS Applied Nano Materials 2~(1) (2018) 202--213.
	
	\bibitem{ABDULLAH2021110095}
	N.~R. Abdullah, H.~O. Rashid, C.-S. Tang, A.~Manolescu, V.~Gudmundsson,
	\href{https://www.sciencedirect.com/science/article/pii/S002236972100161X}{Role
		of interlayer spacing on electronic, thermal and optical properties of
		BN-codoped bilayer graphene: Influence of the interlayer and the induced
		dipole-dipole interactions}, Journal of Physics and Chemistry of Solids 155
	(2021) 110095.
	\newblock \href {https://doi.org/https://doi.org/10.1016/j.jpcs.2021.110095}
	{\path{doi:https://doi.org/10.1016/j.jpcs.2021.110095}}.
	\newline\urlprefix\url{https://www.sciencedirect.com/science/article/pii/S002236972100161X}
	
	\bibitem{das2021electronic}
	A.~Das, R.~Yadav, Electronic and vibrational properties of pristine and Cd, Si,
	Zn and Ge-doped InN nanosheet: a first principle study, Structural Chemistry
	32~(1) (2021) 379--386.
	
	\bibitem{ghasemzadeh2018strain}
	F.~Ghasemzadeh, F.~Kanjouri, Strain effect on the electronic properties of
	III-Nitride nanosheets: Ab-initio study, Science China Technological Sciences
	61~(4) (2018) 535--541.
	
	\bibitem{jalilian2016tuning}
	J.~Jalilian, M.~Naseri, S.~Safari, M.~Zarei, Tuning of the electronic and
	optical properties of single-layer Indium Nitride by strain and stress,
	Physica E: Low-dimensional Systems and Nanostructures 83 (2016) 372--377.
	
	\bibitem{postorino2020strain}
	S.~Postorino, D.~Grassano, M.~D’Alessandro, A.~Pianetti, O.~Pulci,
	M.~Palummo, Strain-induced effects on the electronic properties of 2D
	materials, Nanomaterials and Nanotechnology 10 (2020) 1847980420902569.
	
	\bibitem{hoat2020reducing}
	D.~Hoat, M.~Naseri, R.~Ponce-P{\'e}rez, N.~N. Hieu, T.~V. Vu, J.~Rivas-Silva,
	G.~H. Cocoletzi, Reducing the electronic band gap of BN monolayer by
	coexistence of P (as)-doping and external electric field, Superlattices and
	Microstructures 137 (2020) 106357.
	
	\bibitem{ABDULLAH2021114644}
	N.~R. Abdullah, M.~T. Kareem, H.~O. Rashid, A.~Manolescu, V.~Gudmundsson,
	\href{https://www.sciencedirect.com/science/article/pii/S1386947721000266}{Spin-polarised
		DFT modeling of electronic, magnetic, thermal and optical properties of
		silicene doped with transition metals}, Physica E: Low-dimensional Systems
	and Nanostructures 129 (2021) 114644.
	\newblock \href {https://doi.org/https://doi.org/10.1016/j.physe.2021.114644}
	{\path{doi:https://doi.org/10.1016/j.physe.2021.114644}}.
	\newline\urlprefix\url{https://www.sciencedirect.com/science/article/pii/S1386947721000266}
	
	\bibitem{aggoune2020structural}
	W.~Aggoune, C.~Cocchi, D.~Nabok, K.~Rezouali, M.~A. Belkhir, C.~Draxl,
	Structural, electronic, and optical properties of periodic graphene/h-BN van
	der waals heterostructures, Physical Review Materials 4~(8) (2020) 084001.
	
	\bibitem{jalilian2016buckling}
	J.~Jalilian, M.~Safari, S.~Naderizadeh, Buckling effects on electronic and
	optical properties of BeO monolayer: First principles study, Computational
	Materials Science 117 (2016) 120--126.
	
	\bibitem{doi:10.1063/1.4979507}
	J.~Wu, Y.~Yang, H.~Gao, Y.~Qi, J.~Zhang, Z.~Qiao, W.~Ren,
	\href{https://doi.org/10.1063/1.4979507}{Electric field effect of GaAs
		monolayer from first principles}, AIP Advances 7~(3) (2017) 035218.
	\newblock \href {http://arxiv.org/abs/https://doi.org/10.1063/1.4979507}
	{\path{arXiv:https://doi.org/10.1063/1.4979507}}, \href
	{https://doi.org/10.1063/1.4979507} {\path{doi:10.1063/1.4979507}}.
	\newline\urlprefix\url{https://doi.org/10.1063/1.4979507}
	
	\bibitem{Giannozzi_2009}
	P.~Giannozzi, S.~Baroni, N.~Bonini, M.~Calandra, R.~Car, C.~Cavazzoni,
	D.~Ceresoli, G.~L. Chiarotti, M.~Cococcioni, I.~Dabo, A.~D. Corso,
	S.~de~Gironcoli, S.~Fabris, G.~Fratesi, R.~Gebauer, U.~Gerstmann,
	C.~Gougoussis, A.~Kokalj, M.~Lazzeri, L.~Martin-Samos, N.~Marzari, F.~Mauri,
	R.~Mazzarello, S.~Paolini, A.~Pasquarello, L.~Paulatto, C.~Sbraccia,
	S.~Scandolo, G.~Sclauzero, A.~P. Seitsonen, A.~Smogunov, P.~Umari, R.~M.
	Wentzcovitch,
	\href{https://doi.org/10.1088%2F0953-8984%2F21%2F39%2F395502}{{QUANTUM}
		{ESPRESSO}: a modular and open-source software project for quantum
		simulations of materials}, Journal of Physics: Condensed Matter 21~(39)
	(2009) 395502.
	\newblock \href {https://doi.org/10.1088/0953-8984/21/39/395502}
	{\path{doi:10.1088/0953-8984/21/39/395502}}.
	\newline\urlprefix\url{https://doi.org/10.1088%2F0953-8984%2F21%2F39%2F395502}
	
	\bibitem{giannozzi2017advanced}
	P.~Giannozzi, O.~Andreussi, T.~Brumme, O.~Bunau, M.~B. Nardelli, M.~Calandra,
	R.~Car, C.~Cavazzoni, D.~Ceresoli, M.~Cococcioni, et~al., Advanced
	capabilities for materials modelling with quantum espresso, Journal of
	Physics: Condensed Matter 29~(46) (2017) 465901.
	
	\bibitem{PhysRevLett.77.3865}
	J.~P. Perdew, K.~Burke, M.~Ernzerhof,
	\href{https://link.aps.org/doi/10.1103/PhysRevLett.77.3865}{Generalized
		gradient approximation made simple}, Phys. Rev. Lett. 77 (1996) 3865--3868.
	\newblock \href {https://doi.org/10.1103/PhysRevLett.77.3865}
	{\path{doi:10.1103/PhysRevLett.77.3865}}.
	\newline\urlprefix\url{https://link.aps.org/doi/10.1103/PhysRevLett.77.3865}
	
	\bibitem{doi:10.1063/1.1926272}
	J.~Paier, R.~Hirschl, M.~Marsman, G.~Kresse,
	\href{https://doi.org/10.1063/1.1926272}{The Perdew–Burke–Ernzerhof
		exchange-correlation functional applied to the G2-1 test set using a
		plane-wave basis set}, The Journal of Chemical Physics 122~(23) (2005)
	234102.
	\newblock \href {http://arxiv.org/abs/https://doi.org/10.1063/1.1926272}
	{\path{arXiv:https://doi.org/10.1063/1.1926272}}, \href
	{https://doi.org/10.1063/1.1926272} {\path{doi:10.1063/1.1926272}}.
	\newline\urlprefix\url{https://doi.org/10.1063/1.1926272}
	
	\bibitem{ABDULLAH2022106835}
	N.~R. Abdullah, B.~J. Abdullah, V.~Gudmundsson,
	\href{https://www.sciencedirect.com/science/article/pii/S1293255822000309}{DFT
		study of tunable electronic, magnetic, thermal, and optical properties of a
		Ga$_2$Si$_6$ monolayer}, Solid State Sciences 125 (2022) 106835.
	\newblock \href
	{https://doi.org/https://doi.org/10.1016/j.solidstatesciences.2022.106835}
	{\path{doi:https://doi.org/10.1016/j.solidstatesciences.2022.106835}}.
	\newline\urlprefix\url{https://www.sciencedirect.com/science/article/pii/S1293255822000309}
	
	\bibitem{ABDULLAH2022114590}
	N.~R. Abdullah, B.~J. Abdullah, H.~O. Rashid, C.-S. Tang, V.~Gudmundsson,
	\href{https://www.sciencedirect.com/science/article/pii/S0038109821003732}{Modulation
		of electronic and thermal proprieties of TaMoS$_2$ by controlling the repulsive
		interaction between Ta dopant atoms}, Solid State Communications 342 (2022)
	114590.
	\newblock \href {https://doi.org/https://doi.org/10.1016/j.ssc.2021.114590}
	{\path{doi:https://doi.org/10.1016/j.ssc.2021.114590}}.
	\newline\urlprefix\url{https://www.sciencedirect.com/science/article/pii/S0038109821003732}
	
	
	\bibitem{Liu_2019}
	X.-F. Liu, Z.-J. Luo, X.~Zhou, J.-M. Wei, Y.~Wang, X.~Guo, B.~Lv, Z.~Ding,
	\href{https://doi.org/10.1088/1674-1056/28/8/086105}{Structural, mechanical,
		and electronic properties of 25 kinds of {III}{\textendash}V binary
		monolayers: A computational study with first-principles calculation}, Chinese
	Physics B 28~(8) (2019) 086105.
	\newblock \href {https://doi.org/10.1088/1674-1056/28/8/086105}
	{\path{doi:10.1088/1674-1056/28/8/086105}}.
	\newline\urlprefix\url{https://doi.org/10.1088/1674-1056/28/8/086105}
	
	\bibitem{gacevic:hal-00632224}
	Z.~Gacevic, P.~Lefebvre, F.~Bertram, G.~Schmidt, P.~Veit, J.~Christen,
	E.~Calleja, \href{https://hal.archives-ouvertes.fr/hal-00632224}{{Growth and
			Characterization of InGaN/GaN Quantum Dots for violet-blue Applications}},
	{9th International Conference on Nitride Semiconductors - ICNS9.}, poster
	(Jul 2011).
	\newline\urlprefix\url{https://hal.archives-ouvertes.fr/hal-00632224}
	
	\bibitem{kaufman1993inorganic}
	G.~B. Kaufman, Inorganic chemistry: principles of structure and reactivity,
	(huheey, james e.; keiter, ellen a.; keiter, richard l.) (1993).
	
	\bibitem{doi:10.1063/1.1564060}
	J.~Heyd, G.~E. Scuseria, M.~Ernzerhof,
	\href{https://doi.org/10.1063/1.1564060}{Hybrid functionals based on a
		screened coulomb potential}, The Journal of Chemical Physics 118~(18) (2003)
	8207--8215.
	\newblock \href {http://arxiv.org/abs/https://doi.org/10.1063/1.1564060}
	{\path{arXiv:https://doi.org/10.1063/1.1564060}}, \href
	{https://doi.org/10.1063/1.1564060} {\path{doi:10.1063/1.1564060}}.
	\newline\urlprefix\url{https://doi.org/10.1063/1.1564060}
	
	\bibitem{abdullah2022enhanced}
	N.~R. Abdullah, B.~J. Abdullah, C.-S. Tang, V.~Gudmundsson, Enhanced
	ultraviolet absorption in BN monolayers caused by tunable buckling, arXiv
	preprint arXiv:2201.00116 (2022).
	
	\bibitem{Ren2012}
	X.~Ren, P.~Rinke, C.~Joas, M.~Scheffler,
	\href{https://doi.org/10.1007/s10853-012-6570-4}{Random-phase approximation
		and its applications in computational chemistry and materials science},
	Journal of Materials Science 47~(21) (2012) 7447--7471.
	\newblock \href {https://doi.org/10.1007/s10853-012-6570-4}
	{\path{doi:10.1007/s10853-012-6570-4}}.
	\newline\urlprefix\url{https://doi.org/10.1007/s10853-012-6570-4}
	
	\bibitem{PhysRev.128.2093}
	D.~R. Penn,
	\href{https://link.aps.org/doi/10.1103/PhysRev.128.2093}{Wave-number-dependent
		dielectric function of semiconductors}, Phys. Rev. 128 (1962) 2093--2097.
	\newblock \href {https://doi.org/10.1103/PhysRev.128.2093}
	{\path{doi:10.1103/PhysRev.128.2093}}.
	\newline\urlprefix\url{https://link.aps.org/doi/10.1103/PhysRev.128.2093}
	
\end{thebibliography}

\end{document}